# Image Inpainting by Hyperbolic Selection of Pixels for Two Dimensional Bicubic Interpolations

Mehran Motmaen
Department of Electrical and Computer Engineering
Isfahan University of Technology
Isfahan, Iran
Email:
motmaen73@gmail.com

Majid Mohrekesh
Department of Electrical and Computer Engineering
Isfahan University of Technology
Isfahan, Iran
Email:
mmohrekesh@yahoo.com

Mojtaba Akbari
Department of Electrical and Computer Engineering
Isfahan University of Technology
Isfahan, Iran
Email:
mojtaba.akbari@ec.iut.ac.ir

Nader Karimi
Department of Electrical and Computer Engineering
Isfahan University of Technology
Isfahan, Iran
Email:
nader.karimi@cc.iut.ac.ir

Shadrokh Samavi
Department of Electrical and Computer Engineering
Isfahan University of Technology
Isfahan, Iran
Email:
samavi96@cc.iut.ac.ir

*Abstract*— **Image inpainting is a restoration process which has numerous applications. Restoring of scanned old images with scratches, or removing objects in images are some of inpainting applications. Different approaches have been used for implementation of inpainting algorithms. Interpolation approaches only consider one direction for this purpose. In this paper we present a new perspective to image inpainting. We consider multiple directions and apply both one-dimensional and two-dimensional bicubic interpolations. Neighboring pixels are selected in a hyperbolic formation to better preserve corner pixels. We compare our work with recent inpainting approaches to show our superior results.**

## I. Introduction

Image inpainting refers to filling missing places based on neighbor regions. Inpainting algorithms works based on human perception of image and also image saliency. These algorithms find degraded places using degradation mask. One of the important goals of any inpainting algorithm is to create artificial textures similar to neighbor pixels.

There are different methods in this field based on different requirements and image features. These methods could be categorized in one dimensional or two dimensional image features. Different libraries are available for image inpainting. For example, [1] is an open source library for image restoration that can be used as image inpainting tool. Library of [1] automatically finds degraded pixels and fills them with an approximation of neighbor pixels. The application in [2] for image inpainting is available online for image enhancement and is especially appropriate for old images.

Some algorithms use linear structure methods for image restoration and inpainting. Linear structure method uses partial differential equations for removing objects in image. The methods proposed in [3], [4], [5] and [6] are the leading methods for inpainting that use some sort of linear structure. In previous works, several researchers have considered texture synthesis as a way to fill large image regions with pure textures – repetitive two-dimensional textural patterns with moderate stochasticity [7]. The work in [8] is used for inpainting in order to satisfy the neighborhood coherency of missed patches.

Method proposed in [6], [9] and [10] uses exemplar based inpainting method. Exemplar based inpainting generates new textures by sampling and copying colors from the source region. This exemplar based methods have difficulties in filling holes of real world scenes.

In this paper we circumvent the deficiencies of other methods by using an aggregation mechanism on the results of 1D and 2D bicubic algorithms. Structure of the paper is as follows: in section 2 we present the proposed method. In section 3 experimented results are presented. Section 4 belongs to concluding remarks.

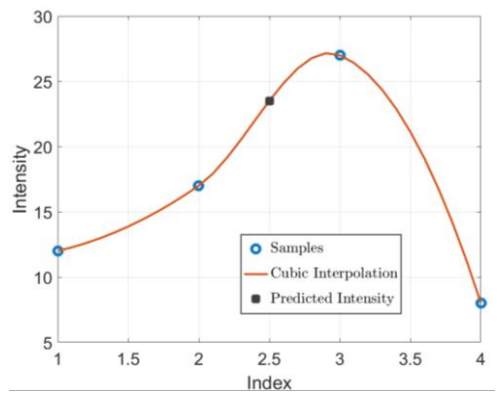

Figure 1.  One dimensional interpolation using four known points.

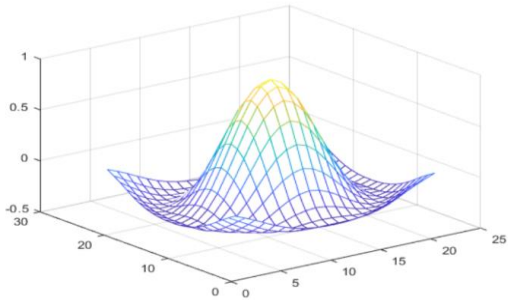

Figure 2. Two dimensional interpolation by fitting a surface.

## II. PROPOSED METHOD

Our proposed method takes advantage of bicubic interpolation and aggregates different results of various such interpolations to produce a final result for a missing pixel. Bicubic interpolation is a graphical method for increasing or decreasing the number of pixels in digital images. Digital cameras use this method for zoom in or zoom out when the desired resolution is different from that of their sensor. The quality of final image depends on the precision and efficiency of the interpolation method.

Bicubic interpolation is a complex method of interpolation that produces better edges and causes less blockiness. Fig. 1 demonstrates a one dimensional interpolation of four instances in a curve. The four samples produce a curve and a single predicted intensity is produced on the curve. Fig. 2 shows a 2-dimensional output of bicubic interpolation to produce a denser mesh from a low number of instances. The resulted mesh is 21×21 made of an 11×11 original grid function. The extra samples are generated by interpolation.

Fig. 3 is an example of degraded image and the mask of degradation. The mask is a binary map of degraded pixels. Fig. 4 demonstrates the block diagram of our proposed method. The input data of the algorithm consist of one degraded image and the mask of degradation which addresses the coordinates of degraded pixels. The mask is a logical map of degraded pixels. Block (1) produces a patch of 16 pixels around the missing target. The 16 pixels that are considered for a missing pixel are demonstrated in Fig. 5. Block (1) of the algorithm sends the mentioned 16 pixels to the two following blocks. Block (2) produces four lines, each consisting of four pixels. These lines consist of one horizontal, one vertical, and two diagonals. The output of Block (2) is sent to Block (3). This block assumes each line as a five-pixel line with a missing center and predicts an intensity value for the center pixel using 1-D bicubic interpolation. Hence, there will be 4 predictions for the center pixel. The missing pixel is the intersection point of all four lines.

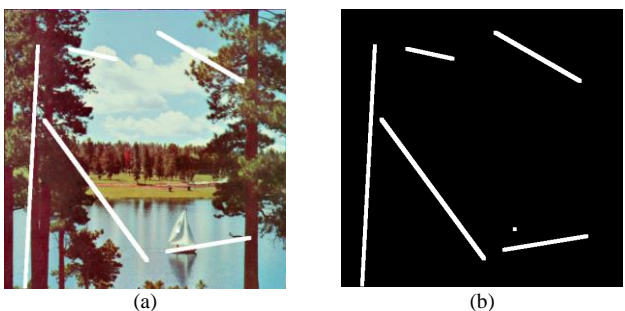

Figure 3. (a) Degraded image and (b) the mask

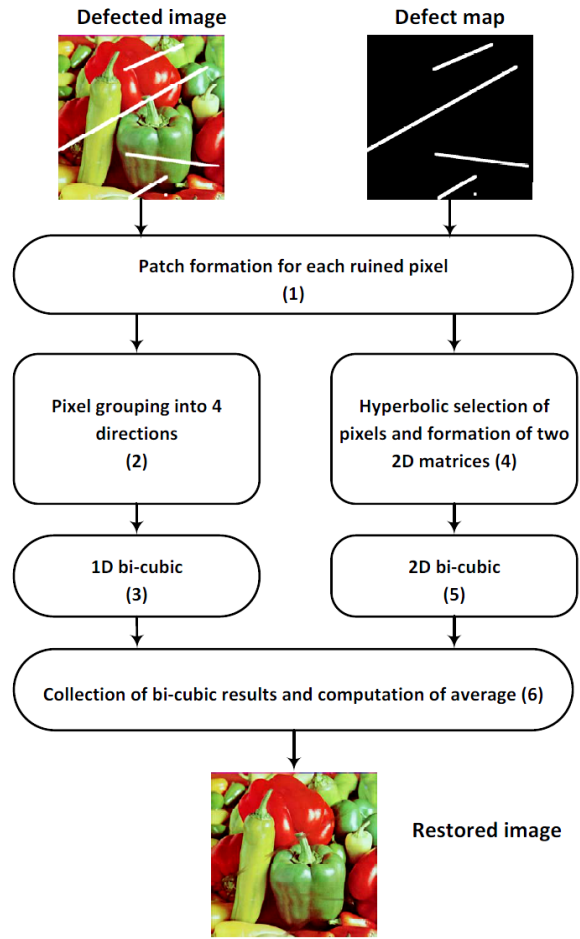

Figure 4. Block diagram of algorithm

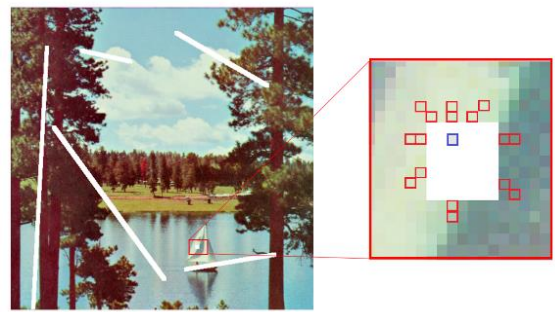

Figure 5. Blue target and the 16 neighbor red pixels

In the right path the block diagram of Fig. 4, below Block (1), there is Block (4), which forms two 12-element matrices as is demonstrated in Fig. 6. In Fig. 6 a 5×5 patch is demonstrated around a single missing pixel. Each pixel of the patch is addressed by an integer number between 1 and 25. Fig. 6(a) and Fig. 6(b) show the selection policy that is used by Block (4). Two 12-pixel formations of the entire 25 pixels use a hyperbolic approach for selection of pixels from the patch. Fig. 6(a) and Fig. 6(b) show the horizontal and vertical hyperbolas respectively. The center of symmetry in both selections are placed on the missing pixel. This hyperbolic selection enhances the corners and prevents the interpolation algorithm to weaken such corners.

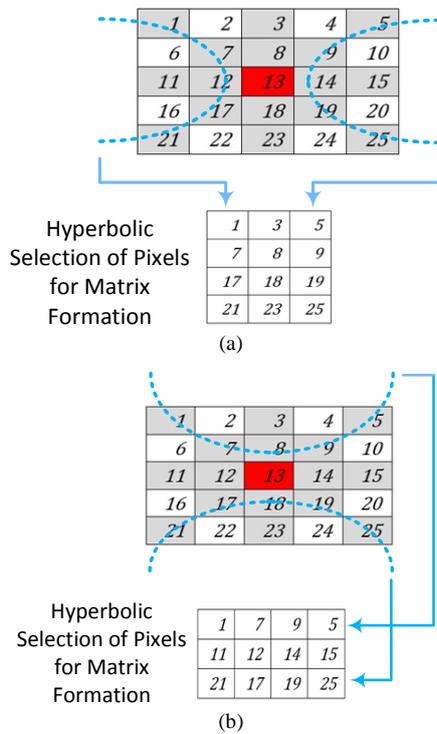

Figure 6. Matrix formation for 2-D bicubic interpolation by hyperbolic selection of pixels. Patch of 25 neighboring pixels is shown (a) 4×3 selection, (b) 3×4 selection

Block (5) performs 2-D bicubic interpolation on both 12 element paths and produces two new matrices, a 7×5 matrix as the interpolation output of the 4×3 output of Block (4) and a 5×7 for the result of performing interpolation on the 3×4 matrix. The two resulted matrices have two centers as two different predictions for the missing target.

Predictions are performed in two distinct prediction modules and the results are aggregated to produce the final intensity value. One module tries to predict the result in one dimensional directions and the other interpolates in two dimensional patches of intensities. Both modules predict the value of intensity using bicubic interpolation.

Block (2) groups the 16 pixels to four lines of four pixels of any direction and Block (3) produces one value of intensity for each direction and results in four values for the target. Then the most different predicted value among these four values are replaced with the average of the other three values. The one dimensional prediction result is the four-member set of intensities in the output of Block (3).

Blocks (4) and (5) are the building blocks of 2-dimensional prediction in our proposed inpainting method. Block (5) gets the 16 pixels from Block (1) and produces a pair of one 3×4 and one 4×3 matrix. The two sets are then delivered to Block (5) for 2-dimensional interpolation. There a new set of 5×7 and 7×5 matrices are created from the input set of 3×4 and 4×3 matrices respectively. The central point of these new sets of matrices are two new predicted values for the target pixel.

These two intensity values, with the four values from one-dimensional predictions, produce six values for Block (6) to produce the final result by averaging this set of values.

TABLE I. COMPARISON OF PROPOSED METHOD WITH PROPOSED INPAINTING METHOD IN [7] AND [10]

| IMAGE | Criterion | Korman [8] | Criminisi [7] | Proposed Method |
|---|---|---|---|---|
|  | **PSNR(dB)** | 29.62 | 41.01 | **46.54** |
|  | **SSIM** | 0.995 | 0.999 | **0.999** |
|  | **PSNR(dB)** | 31.04 | 32.73 | **37.17** |
|  | **SSIM** | 0.987 | 0.991 | **0.995** |
|  | **PSNR(dB)** | 26.78 | 27.68 | **30.82** |
|  | **SSIM** | 0.968 | 0.972 | **0.981** |
|  | **PSNR(dB)** | 27.29 | 29.14 | **30.01** |
|  | **SSIM** | 0.966 | 0.973 | **0.972** |

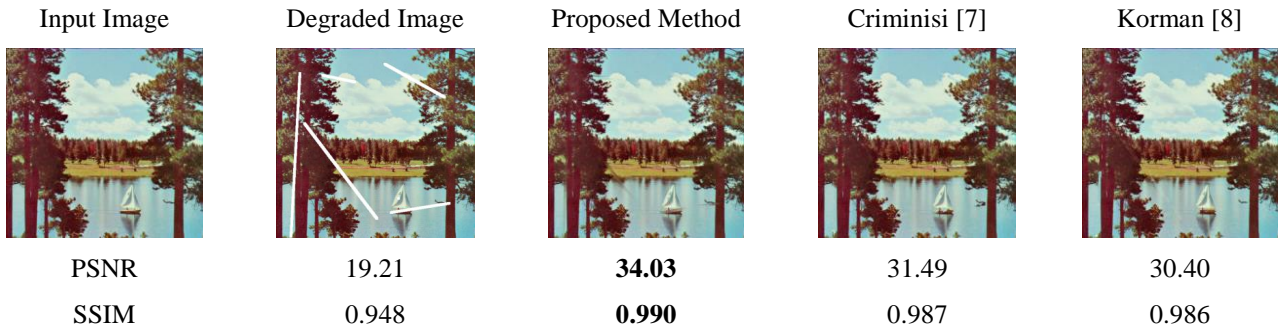

Figure 7. Comparison of Proposed Method with the Method in [7] and [10]

### III. EXPERIMENTAL RESULTS

Fig. 7 shows a typical output of our proposed method in comparison with the method proposed in [7] and [8]. Fig. 8 shows change of PSNR values with respect to the number of defective lines that are present in the image. It can be seen that increasing the number of lines would increase the total number of missing pixels. When a missing pixel is restored by interpolation, the estimated value would be close but different from the original value. This cause a degradation in the PSNR value. But it can be seen that our algorithm results in lower degradation of the restored image. This means that our algorithm can predict the missing pixel better than the other two methods. Fig. 9 shows changes that occur in the obtained PSNR with respect to the width of defective lines for the proposed method as compared to the methods of [7] and [8]. In this experiment initially two defect lines are introduced with the width of one pixel. Then widths of these lines are increased by one pixel at each iteration, all the way up to width of 15 pixels. At each iteration the missing pixels are restored using our proposed method and the methods of [7] and [10]. Then the PSNR value for each method is calculated. The plot of Fig. 9 shows that as the widths of the lines are increased the overall PSNR value would decrease. But it can be seen that our results for any line width has higher PSNR values. Furthermore, we compared results of our proposed method with that of similar works when we have equal amount of degradations. Table I shows comparison of the proposed method with the methods of [7] and [8]. Our proposed method has better quality assessment results in terms of PSNR and SSIM.

### IV. CONCLUSION

In this paper we proposed a novel method of inpainting for the degradations that are in the form of lines. These types of deforms are very probable, especially in the case of scanning of old photographs. Our method is based on one dimensional and two dimensional bicubic interpolations. When selecting neighboring pixels for interpolation, we consider pixels that form hyperbolic paths. By doing this type of selective interpolation we preserve corners in four directions in most cases. The experimental results prove better performance of our method in most cases as are shown by graphs of "Fig. 8" and Fig. 9.

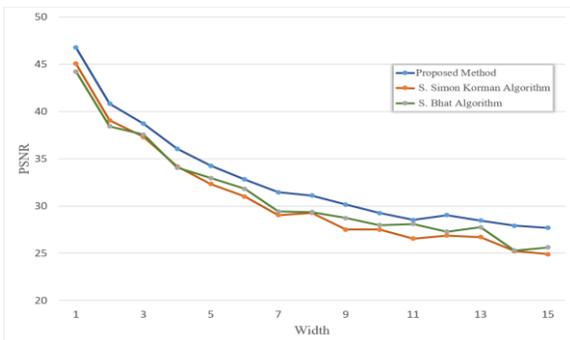

Figure 9. Variation of PSNR for different widths of defective lines.

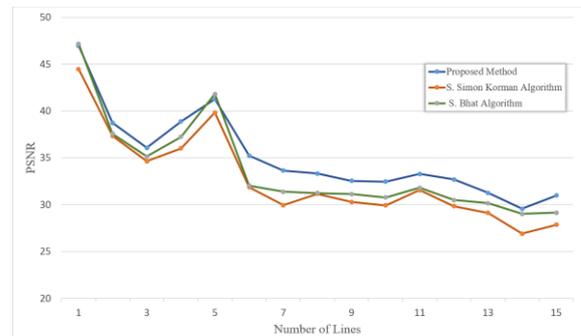

Figure 8. Variation of PSNR for different number of defective lines